\documentclass[aps,prc,amsfonts,showpacs,superscriptaddress,groupedaddress,nofootinbib,floatfix]{revtex4}  
\pdfoutput=1
\usepackage{graphics}  

\usepackage{dcolumn}   
\usepackage{bm}        
\usepackage{amssymb}   
\usepackage{color}

\usepackage{psfrag}
\usepackage{epsfig}
\usepackage{epsf}
\usepackage{float}
\usepackage{slashed}

\newcommand{\beq}{\begin{equation}}
\newcommand{\eeq}{\end{equation}}
\newcommand{\beqa}{\begin{eqnarray}}
\newcommand{\eeqa}{\end{eqnarray}}
\newcommand{\nn}{\nonumber \\ }

\newcommand{\fet}[1]{\mbox{\boldmath $#1$}}

\begin{document}

\title{Few- and many-nucleon systems with semilocal coordinate-space
  regularized chiral two- and three-body forces}




\author{E.~Epelbaum}
\affiliation{Institut f\"ur Theoretische Physik II, Ruhr-Universit\"at
  Bochum, D-44780 Bochum, Germany}


\author{J.~Golak}
\affiliation{M. Smoluchowski Institute of Physics, Jagiellonian
University,  PL-30348 Krak\'ow, Poland}

\author{K.~Hebeler}
\affiliation{Institut f\"ur Kernphysik, Technische Universit\"at 
Darmstadt, 64289 Darmstadt, Germany}

\author{T.~H\"uther}
\affiliation{Institut f\"ur Kernphysik, Technische Universit\"at 
Darmstadt, 64289 Darmstadt, Germany}

\author{H.~Kamada}
\affiliation{Department of Physics, Faculty of Engineering,
Kyushu Institute of Technology, Kitakyushu 804-8550, Japan}

\author{H.~Krebs}
\affiliation{Institut f\"ur Theoretische Physik II, Ruhr-Universit\"at
  Bochum, D-44780 Bochum, Germany}


\author{P.~Maris}
\affiliation{Department of Physics and Astronomy, Iowa State
  University, Ames, Iowa 50011, USA}

\author{Ulf-G.~Mei{\ss}ner}
\affiliation{Helmholtz-Institut~f\"{u}r~Strahlen-~und~Kernphysik~and~Bethe~Center~for~Theoretical~Physics,
~Universit\"{a}t~Bonn,~D-53115~Bonn,~Germany}
\affiliation{Institut f\"ur Kernphysik, Institute for Advanced Simulation 
and J\"ulich Center for Hadron Physics, Forschungszentrum J\"ulich, 
D-52425 J\"ulich, Germany}
\affiliation{JARA~-~High~Performance~Computing,~Forschungszentrum~J\"{u}lich,~D-52425~J\"{u}lich,~Germany}


\author{A.~Nogga}
\affiliation{Institut f\"ur Kernphysik, Institute for Advanced Simulation 
and J\"ulich Center for Hadron Physics, Forschungszentrum J\"ulich, 
D-52425 J\"ulich, Germany}

\author{R.~Roth}
\affiliation{Institut f\"ur Kernphysik, Technische Universit\"at 
Darmstadt, 64289 Darmstadt, Germany}

\author{R.~Skibi\'nski}
\affiliation{M. Smoluchowski Institute of Physics, Jagiellonian
University,  PL-30348 Krak\'ow, Poland}

\author{K.~Topolnicki}
\affiliation{M. Smoluchowski Institute of Physics, Jagiellonian
University,  PL-30348 Krak\'ow, Poland}

\author{J.P.~Vary}
\affiliation{Department of Physics and Astronomy, Iowa State
  University, Ames, Iowa 50011, USA}

\author{K.~Vobig}
\affiliation{Institut f\"ur Kernphysik, Technische Universit\"at 
Darmstadt, 64289 Darmstadt, Germany}

\author{H.~Wita{\l}a}
\affiliation{M. Smoluchowski Institute of Physics, Jagiellonian
University,  PL-30348 Krak\'ow, Poland}

\collaboration{LENPIC Collaboration}

\date{\today}

\begin{abstract}
We present a \emph{complete} calculation of nucleon-deuteron
scattering as well as ground and low-lying excited states of light
nuclei in the mass range A=3-16 up through next-to-next-to-leading order 
in chiral effective field
theory using semilocal coordinate-space regularized two- and
three-nucleon forces. It is shown that both of the low-energy
constants entering the three-nucleon force at this order can be reliably
determined from the triton binding energy and the differential cross
section minimum in elastic nucleon-deuteron scattering. The inclusion
of the three-nucleon force is found to improve the agreement with the
data for most of the considered observables. 
\end{abstract}



\maketitle

\section{Introduction}

Chiral effective field theory (EFT) offers a convenient and powerful
framework to analyze low-energy properties of few- and many-body
nuclear systems in harmony with the symmetries (and their breaking
pattern) of QCD, see
\cite{Epelbaum:2008ga,Machleidt:2011zz,Epelbaum:2012vx} for review
articles. In recent years, the chiral expansion of the
two-nucleon (NN) force, the dominant part of the nuclear Hamiltonian, has
been pushed to fifth order (N$^4$LO)
\cite{Entem:2014msa,Epelbaum:2014sza,Entem:2017gor,Reinert:2017usi}
and even beyond \cite{Entem:2015xwa}. 
The available versions of the N$^4$LO potentials differ from each
other, among other things, in the functional form of the regulator
function: while the interactions of Ref.~\cite{Entem:2017gor} are regularized with
a nonlocal cutoff, local regularization in coordinate  (momentum)
space is employed for pion-exchange contributions in Ref.~\cite{Epelbaum:2014sza} 
(Ref.~\cite{Reinert:2017usi}). As demonstrated in Refs.~\cite{Epelbaum:2014efa,Reinert:2017usi},
the employed types of local regulators do, per construction, not
affect the long-range part of the interaction thus generating a smaller
amount of finite-cutoff artifacts. For a related discussion of
regulator artifacts in uniform matter see
Ref.~\cite{Dyhdalo:2016ygz}. The resulting N$^4$LO$^+$ potentials of
Ref.~\cite{Reinert:2017usi} lead to the description of the 2013
Granada data base \cite{Perez:2013mwa} for
neutron-proton and proton-proton scattering below $E_{\rm lab} =
300$~MeV which is comparable to or even better than that based on the
phenomenological  high-precision potentials such as the AV18
\cite{AV18}, CDBonn \cite{cdb}, Nijm1 and 
Nijm2 \cite{nijm} models, featuring at the same time a much smaller
number of adjustable parameters.  We also mention recent efforts towards
constructing the NN
\cite{Piarulli:2014bda,Piarulli:2016vel,Ekstrom:2017koy} and
three-nucleon \cite{Epelbaum:2007sq,Krebs:2018jkc} potentials using the heavy-baryon formulation
of chiral EFT with explicit $\Delta(1232)$ degrees of freedom. 

In Refs.~\cite{Binder:2015mbz,Maris:2016wrd,Binder:2018pgl}, we have
applied the semilocal coordinate-space regularized (SCS) chiral NN
potentials of Refs.~\cite{Epelbaum:2014efa,Epelbaum:2014sza} to
analyze nucleon-deuteron (Nd) scattering  along with selected
properties of light- and medium-mass nuclei. For similar studies of
nuclear matter properties, 
selected electroweak processes and nucleon-deuteron radiative capture
reactions see Refs.~\cite{Hu:2016nkw}, \cite{Skibinski:2016dve} and \cite{Skibinski:2017vqs}, respectively. 
All these calculations are based on the NN forces only and thus can
only be regarded as complete at leading (LO) and next-to-leading
orders (NLO) in the chiral expansion. In fact, our main motivation in
these studies was to analyze the convergence pattern of chiral EFT,
estimate the achievable accuracy at various orders and identify
promising observables to look for three-nucleon force (3NF) effects and/or 
meson-exchange-current contributions. To estimate the truncation error
of the chiral expansion, we followed the algorithm formulated in
Ref.~\cite{Epelbaum:2014efa} and modified appropriately to account for missing 3NFs
and meson-exchange currents. For the interpretation, validation  and further
developments of this approach to uncertainty quantification in a Bayesian framework see
Refs.~\cite{Furnstahl:2015rha,Melendez:2017phj}, while the robustness
of this method and possible alternatives are discussed in
Ref.~\cite{Binder:2018pgl}.  One important outcome of these studies
is the observation that many Nd scattering observables
at intermediate energies as well as the energies and radii of light
and medium-mass nuclei calculated with NN forces only show significant
deviations from experimental data, whose magnitude matches well with
the expected size of 3NF contributions in the Weinberg power counting
scheme. 

In this paper we perform, for the first time, \emph{complete}
calculations of few- and many-nucleon systems at third order of the
chiral expansion, i.e.~at N$^2$LO, utilizing semilocal coordinate-space
regulators~\cite{Epelbaum:2014efa,Epelbaum:2014sza,Binder:2015mbz,Binder:2018pgl}. 
We explore different ways to fix the low-energy constants (LECs) $c_D$
and $c_E$ in the three-nucleon sector and show that they can be
reliably determined from the $^3$H binding energy and the differential
cross section minimum in elastic Nd scattering at
intermediate energies. This allows us to make parameter-free
predictions for $A>3$ systems. We provide a comparison of the complete
N$^2$LO results with results at LO and NLO and estimate truncation
errors.  More details regarding the calculations will be presented in separate
publications~\cite{Maris:inprep2018} for $p$-shell nuclei and
\cite{fut3ncont} for Nd scattering. 

Our paper is organized as follows. In section \ref{cdandce} we specify
the regularized expressions of the chiral 3NF at N$^2$LO and discuss
the determination of the LECs $c_D$ and $c_E$. Section \ref{ndscat} is
devoted to Nd elastic scattering, while our predictions for ground
state and excitation energies for $p$-shell nuclei are reported  in
sections \ref{sec:nuclei} and \ref{sec:nuclei2},
respectively. Finally, the main results of our study are summarized in
section \ref{secsummary}.

\section{Determination of $c_D$ and $c_E$}
\label{cdandce}

The N$^2$LO three-nucleon force in momentum space is given by 
\beqa
\label{leading}
V^{\rm 3N} &=& \frac{g_A^2}{8 F_\pi^4}\; 
\frac{\vec \sigma_1 \cdot \vec q_1  \; \vec \sigma_3 \cdot \vec q_3 }{[
  q_1^2 + M_\pi^2] \, [q_3^2 + M_\pi^2]} \;\Big[ \fet \tau_1
  \cdot \fet \tau_3  \, \left( - 4 c_1 M_\pi^2 + 2 c_3 \, \vec q_1 \cdot \vec
    q_3 \right)  
+  c_4 \fet \tau_1
  \times \fet \tau_3  \cdot \fet \tau_2  \; \vec q_1 \times \vec q_3 
\cdot \vec \sigma_2  \Big]  \nn
&-& \frac{g_A \, D}{8 F_\pi^2}\;  
\frac{\vec \sigma_3 \cdot \vec q_3 }{q_3^2 + M_\pi^2} \; 
\fet \tau_1 \cdot \fet \tau_3 \; \vec \sigma_1 \cdot \vec q_3 \; + \; 
\frac{1}{2} E\, \fet \tau_1 \cdot \fet \tau_2\; + \; \mbox{5 permutations}\,, 
\eeqa
where the subscripts refer to the nucleon labels and $\vec q_{i} = \vec p_i \,
' - \vec p_i$,  with $\vec p_i \, '$
and $\vec p_i$ being the final and initial momenta of the nucleon $i$. 
Further, $q_i \equiv | \vec q_i |$, $\sigma_i$  and $\fet \tau_i$ are the Pauli
spin and isospin matrices, respectively, 
$c_i$, $D$ and $E$ denote the corresponding 
LECs while $g_A$ and $F_\pi$ refer to the
nucleon axial coupling and pion decay constant. 
Throughout this work, we use the same values for the
subleading pion-nucleon LECs $c_i$ 
as employed in the NN forces of Ref.~\cite{Epelbaum:2014efa}. 
These are compatible with the recent determinations from the
Roy-Steiner analysis \cite{Hoferichter:2015tha}. 
We also apply a consistent regularization procedure. Specifically,
regularization of the $2\pi$-exchange 3NF is carried out by Fourier
transforming the expressions into coordinate space, see Eq.~(2.11) of
Ref.~\cite{Bernard:2007sp}, and subsequently multiplying them with the
regulator functions used in Ref.~\cite{Epelbaum:2014efa}: 
\beq
V^{3N}_{2 \pi} (\vec r_{12}, \, \vec r_{32}) \; \; \longrightarrow \; \; 
V^{3N}_{2 \pi} (\vec r_{12}, \, \vec r_{32}) \, \bigg[ 1 - \exp \bigg(
-\frac{r_{12}^2}{R^2} \bigg) \bigg]^6\, \bigg[ 1 - \exp \bigg(
-\frac{r_{32}^2}{R^2} \bigg) \bigg]^6\,.
\eeq 
Here, $\vec r_{ij}$ denotes the relative distance between the nucleons
$i$ and $j$. For the one-pion-exchange-contact 3NF term proportional to the LEC
$D$ in Eq.~(\ref{leading}), a similar procedure is employed to regularize the singular
behavior with respect to the momentum transfer $\vec q_3$. In
addition, following Ref.~\cite{Epelbaum:2014efa}, the contact interaction between the
nucleons $1$ and $2$ is regularized by multiplying the momentum-space
matrix elements with a nonlocal Gaussian regulator $\exp(- (p_{12}^2 +
  {p_{12}'} ^2)/\Lambda^2 )$, where $\vec p_{12} = (\vec p_1 - \vec
  p_2 )/2$, ${\vec p}_{12}^{\, \prime} = (\vec p_1^{\, \prime} - \vec
  p_2^{\, \prime} )/2$ and $\Lambda = 2 R^{-1}$. Finally, for the purely
  contact interaction proportional to the LEC $E$, we apply a nonlocal
  regulator in momentum space 
\beq
V^{3N}_{\rm cont}  \; \; \longrightarrow \; \; V^{3N}_{\rm cont}  \,
\exp \bigg( -\frac{4 p_{12}^2 + 3 k_3^2}{4 \Lambda^2} \bigg) \,
\exp \bigg( -\frac{4 {p_{12} '}^2 + 3 {k_3'}^2}{4 \Lambda^2} \bigg) \,,
\eeq
where $\vec k_3 = 2 (\vec p_3 - (\vec p_1 + \vec p_2)/2)/3$ and $\vec
k_3^{\, \prime} = 2 (\vec p_3^{\, \prime} - (\vec p_1^{\, \prime} + \vec p_2^{\, \prime})/2)/3$ are
the corresponding Jacobi momenta. The numerical implementation of
the regularization in the partial wave basis will be detailed in a
separate publication. We have verified the correctness of the
implementation by comparing two independent calculations of matrix
elements of the 3NF. 

\begin{figure}
  \includegraphics[width=0.65\columnwidth]{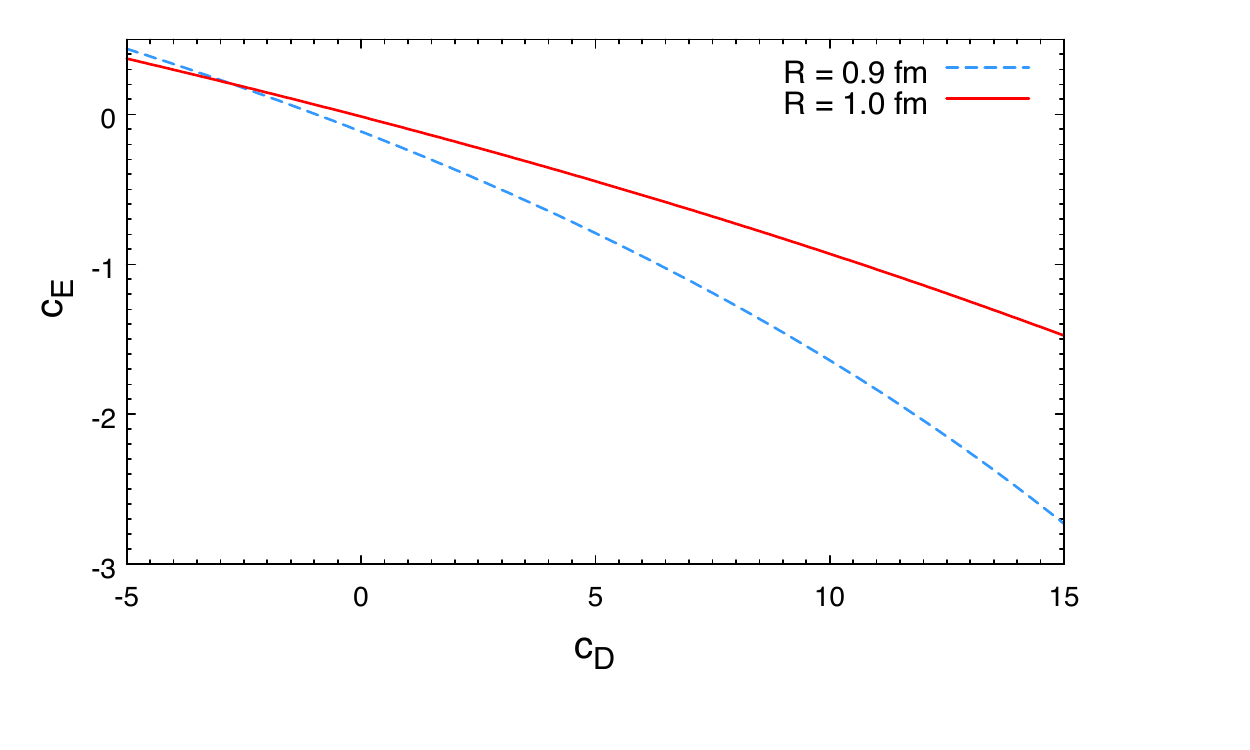}
  \caption{\label{Fig:cDcE} (Color online)
    Correlation between the LECs $c_D$ and $c_E$ induced by the
    requirement to reproduce the $^3$H binding energy for the cutoff
    choices of $R=0.9$~fm (blue dashed line) and $R=1.0$~fm (red solid
    line).}
\end{figure}

The three-nucleon force at N$^2$LO involves two 
LECs which govern the strength of the one-pion-exchange-contact and purely
contact 3NF contributions and cannot be fixed from
nucleon-nucleon scattering. Here and in what follows,
we use the notation of Ref.~\cite{Epelbaum:2002vt} and express these LECs in terms of the
dimensionless parameters $c_D$ and $c_E$ via
\beq
D = \frac{c_D}{F_\pi^2 \Lambda_\chi}, \quad \quad E = \frac{c_E}{F_\pi^4 \Lambda_\chi}\,,
\eeq
employing the value of $\Lambda_\chi = 700$~MeV$\, \simeq M_\rho$ for the
chiral-symmetry breaking scale. The determination of $c_D$
and $c_E$ requires at least two few- or many-nucleon low-energy observables. 
In this analysis we utilize a commonly adopted practice 
\cite{Epelbaum:2002vt,Nogga:2005hp,Navratil:2007we,Gazit:2008ma,Piarulli:2017dwd} and 
regard the $^3$H binding energy as one
such observable. Employing this constraint establishes a relation between
the two LECs as visualized in Fig.~\ref{Fig:cDcE} for the regulator choices
of $R=0.9$~fm and $R=1.0$~fm, which 
leaves us with a single yet undetermined parameter $c_D$.  Notice that
when calculating the $^3$H binding energy to determine $c_E$ as
function of $c_D$, we have taken into account
the electromagnetic interaction between two neutrons as implemented in
the AV18 potential \cite{AV18}.  On the other hand, the results
presented in sections \ref{sec:nuclei} and \ref{sec:nuclei2} are based
on the point-Coulomb interaction only. This small
inconsistency is irrelevant at the accuracy level of our study.

A wide range of observables has been considered in the literature to constrain the
remaining LEC. These  include the neutron-deuteron doublet scattering
length $^2 a$
\cite{Epelbaum:2002vt,Piarulli:2017dwd}, 
triton beta decay \cite{Gazit:2008ma}, the $^4$He binding 
energy \cite{Nogga:2005hp}, the point-proton radii of
$^3$H and/or $^4$He and selected properties of few-nucleon systems 
\cite{Navratil:2007we,Lynn:2017fxg}. We
also mention the approach of Ref.~\cite{Ekstrom:2015rta} 
to perform a global fit of
LECs entering the two- and three-nucleon forces to NN scattering
data in combination with few- and
many-nucleon observables.  
In this paper we explore several possibilities for fixing $c_D$ based
solely on the nucleon-deuteron (Nd) experimental data. Such a
procedure has an advantage of being insensitive to
the four-nucleon force and exchange currents, which may affect
observables in heavier systems and reactions involving electroweak
probes, and gives us the opportunity to make predictions for nuclei
with $A \ge 4$. Given that we only consider the leading contribution to
the 3NF at N$^2$LO, we do not use Nd polarization observables to
determine the $c_D/c_E$ values and
restrict ourselves to the differential and total cross sections and
$^2 a$. Specifically, the differential cross section for
elastic Nd scattering around its minimum at energies of $E_{N}
\sim 50$~MeV and above\footnote{At low energy, the minimum in the
  differential cross section becomes less pronounced due to the S-wave
dominance, and the sensitivity to the 3NF decreases.} 
is well known to be sensitive
to the 3NF contributions \cite{Gloeckle:1995jg,wit98}. 

\begin{figure}
  \includegraphics[width=\columnwidth]{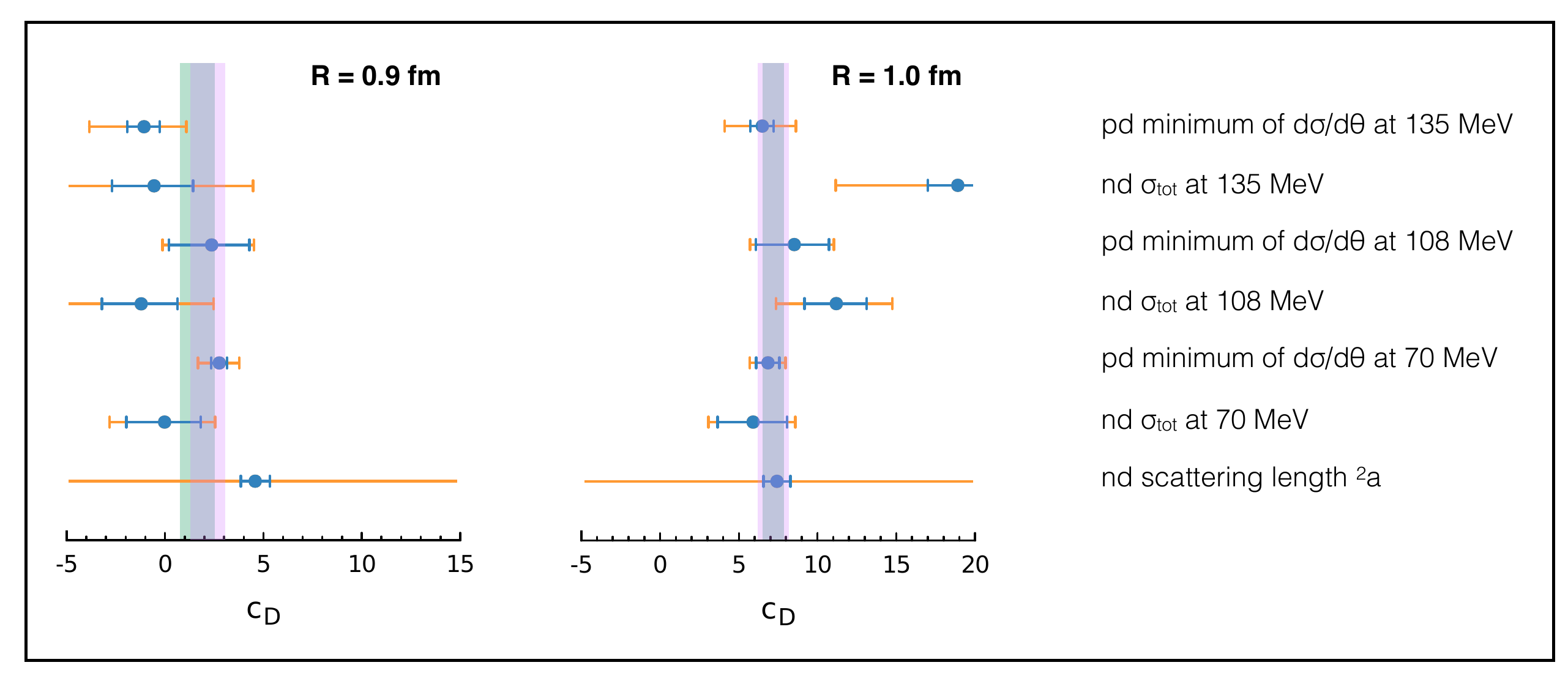}
  \caption{\label{Fig:cD} (Color online)
   Determination of the LEC $c_D$ from the differential cross section
   in elastic pd scattering, total nd cross section and the nd doublet
   scattering length $^2a$ for the cutoff choices of $R=0.9$~fm and 
   $R=1.0$~fm.  The smaller (blue) error bars correspond
   to the experimental uncertainty while the larger (orange) error
   bars also take into account the theoretical uncertainty estimated
   as described in Ref.~\cite{Binder:2015mbz}. The violet (green)
   bands show the results from a combined fit to all observables (to
   observables up to $E_N = 108$~MeV). }
\end{figure} 

In Fig.~\ref{Fig:cD}
we show the constraints on $c_D$ resulting from the reproduction of
the  proton-deuteron differential cross section
data at $E_{N}= 70$ and  $135$~MeV of Ref.~\cite{Sekiguchi:2002sf} and $E_{N}=
108$~MeV of Ref.~\cite{Ermisch:2003zq} for $\theta_{\rm c.m.} \sim
128^\circ$ (using a single experimental point). Notice that there is a discrepancy between the data of 
Ref.~\cite{Sekiguchi:2002sf} and 
the KVI measurement of the differential cross section at
$E_N = 135$~MeV of Ref.~\cite{Ermisch:2005kf}.  
We also show 
constraints emerging from the reproduction of the (derived)
neutron-deuteron total cross section
data of Ref.~\cite{Abfalterer:2001gw} at the same energies and the experimental value
of $^2a = 0.645 \pm 0.008$~fm of Ref.~\cite{Schoen:2003my}.  In
all calculations, the LEC $c_E$ is set to reproduce the $^3$H binding
energy according to the correlation shown in Fig.~\ref{Fig:cDcE}. Notice further that we do not
include the Coulomb interactions in our Nd scattering
calculations. The effect of the Coulomb interaction in the cross
section for the considered kinematics is below the statistical
and systematic uncertainties of our analysis \cite{deltuva2005,Delt}. We further
emphasize that our scattering calculations are carried out including
NN partial waves up to $j_{\rm max} = 5$  and using a standard approximate
treatment of the isospin $T=3/2$ channels, see section \ref{ndscat} for
further information on the partial wave truncation and
 Ref.~\cite{Gloeckle:1995jg} for more
details. This is sufficient to obtain converged results for 
observables under consideration.  
We also neglect isospin $T=3/2$ components of the 3NF when
calculating Nd scattering observables, which are insignificant for the
observables we consider \cite{Witala:2016inj}. 

As shown in Fig.~\ref{Fig:cD}, the strongest constraint on $c_D$
results from the cross section minimum at the lowest considered energy
of $E_N=70$~MeV. While the differential cross section data at
$E_N=135$~MeV have the same statistical and systematic errors,
the significantly larger theoretical uncertainty at this energy leads to a less
precise determination of $c_D$. It is also interesting to see that the
doublet scattering length $^2a$, whose experimental value is
known to a high accuracy of $\sim 1\%$, does not constrain
$c_D$ at N$^2$LO. This is in line with the
known strong correlation between $^2a$ and the $^3$H binding
energy (the so-called Phillips line
\cite{Phillips:1968zze}), see also Ref.~\cite{Gazit:2008ma} for 
 a similar conclusion.  Performing
a $\chi^2$ fit to all considered observables, we obtain  the values of 
$c_D = 1.7 \pm 0.8$ for $R=0.9$~fm and  $c_D = 7.2 \pm 0.7$ for
$R=1.0$~fm. When including the data only up to $108$~MeV, the
  resulting $c_D$ values read  
 $c_D = 2.1 \pm 0.9$ for $R=0.9$~fm and  $c_D = 7.2 \pm 0.9$ for
$R=1.0$~fm. The corresponding $c_E$ values are $c_E =
-0.329^{+0.103}_{-0.106}$ ($c_E =
-0.381^{+0.117}_{-0.122}$) for $R=0.9$~fm and  $c_E =
-0.652 \pm 0.067$ ($c_E =
-0.652^{+0.086}_{-0.087}$) for $R=1.0$~fm using experimental data up
to $135$~MeV (up to $108$~MeV). 

It is important to address the question of robustness of our approach 
to determine the constants $c_D$ and $c_E$. To this end, we 
performed fits to the Nd differential cross section data in 
 a wider range of center-of-mass (c.m.) angles. 
\begin{figure}[tb]   
  \includegraphics[width=0.70\columnwidth]{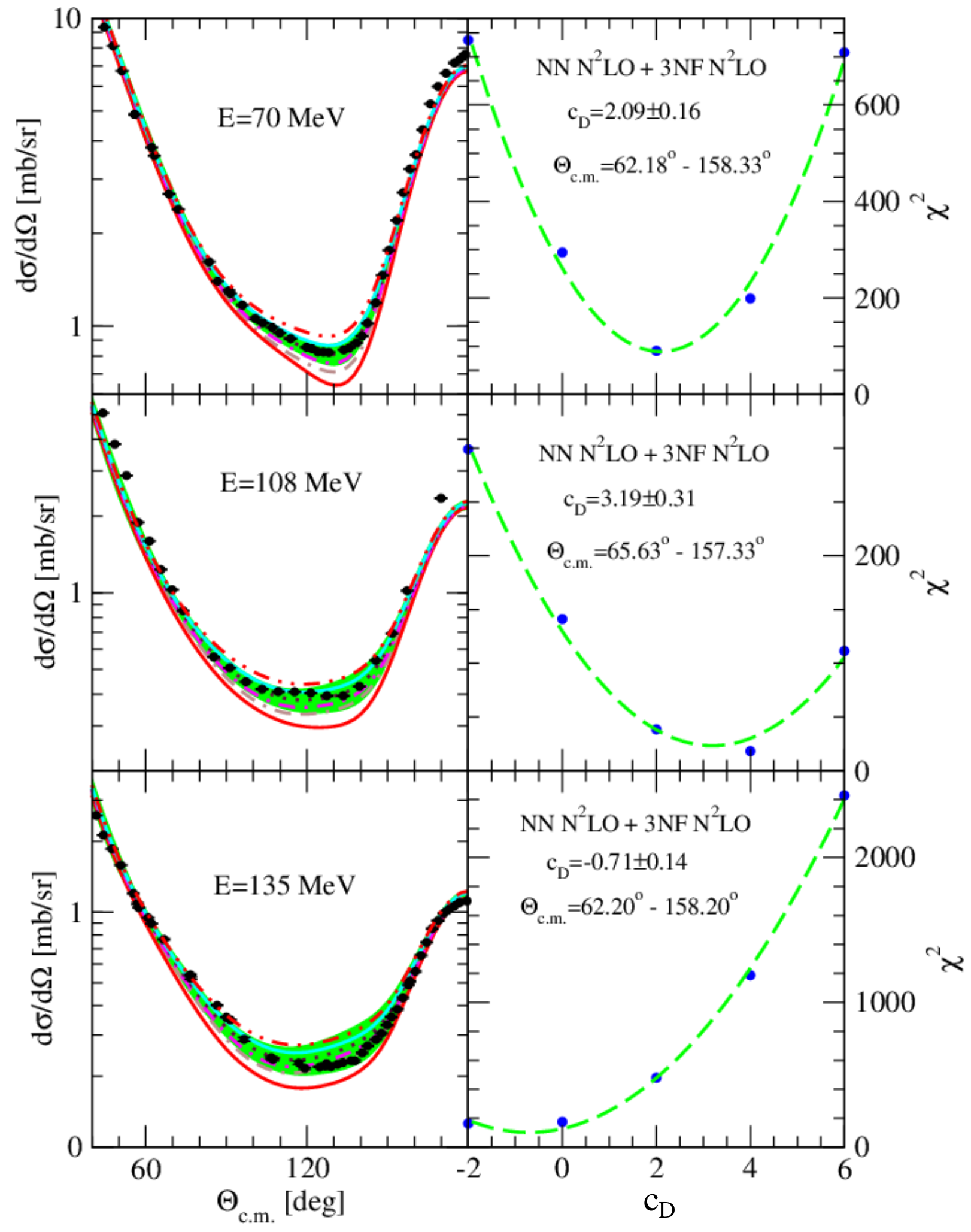}
  \caption{(Color online)  
 The nd elastic scattering cross section at the 
 incoming neutron laboratory
 energies $E=70$, $108$ and $135$~MeV. 
 In the left panel, the solid (red) lines 
 are predictions of 
the N$^2$LO SCS NN potential with the regulator $R=0.9$~fm. Combining this 
NN potential with the N$^2$LO 3NF using five different ($c_D,c_E$) 
combinations  leads to results  
  shown  by the (brown)
 double-dashed-dotted, 
  (magenta) dashed-dotted, 
(maroon)  dotted, (cyan) solid  and (red) 
double-dotted-dashed  lines for $c_D = -2.0$, $0.0$, $2.0$, $4.0$ and
$6.0$, respectively. 
The (green) bands show the estimated theoretical uncertainty of predictions 
 at N$^2$LO with $c_D=2.0$. The 
corresponding $c_E$-values are in all cases taken from the correlation line shown in
Fig.~\ref{Fig:cDcE}.  
The (black) dots depict  
 pd data from Ref.~\cite{Sekiguchi:2002sf} at $E=70$ and $E=135$~MeV and from
 Ref.~\cite{Ermisch:2003zq} at $E=108$~MeV. 
 In the right panel,  the $\chi^2$ fits to 
the experimental data in the indicated angular regions based of on
these five pairs of ($c_D,\; c_E$) values  are shown by 
 dashed (green) lines. The legends in the right pannel provide the
 best fit $c_D$ values to the data at each laboratory energy over the
 indicated angular range.
\label{fig2}}
\end{figure} 
In Fig.~\ref{fig2} we show the resulting description of the data
along with the corresponding $\chi^2$ as a function of $c_D$ for 
the already mentioned pd data at  $E=70$ \cite{Sekiguchi:2002sf}, $108$ \cite{Ermisch:2003zq} and
$135$~MeV \cite{Sekiguchi:2002sf}. The actual
calculations have been performed for $R=0.9$~fm using five different $c_D$ values
namely $c_D = -2.0$, $0.0$, $2.0$, $4.0$ and $6.0$. In all cases, the 
$c_E$-values are taken from the correlation line shown in
Fig.~\ref{Fig:cDcE}.  
The shown $\chi^2$ does not take into account the estimated theoretical
uncertainty of our calculations. Notice further that in all cases, we
have taken into account the systematic errors in addition to the
statistical ones as given in Refs.~\cite{Sekiguchi:2002sf,Ermisch:2005kf}. 
While the resulting $c_D$ values at $70$~MeV and $108$~MeV 
are close to each other and also to the  recommended value of $c_D \sim 2.1$ from the global fit quoted
above,  the fit to the $E=135$~MeV data prefers a value of $c_D \sim -0.7$. However,
taking into account the relatively large theoretical uncertainty at $E=135$~MeV, the
extracted values of $c_D$ at all three energies are still compatible with each other, see
the left graphs of Fig.~\ref{Fig:cD} and left panels of \ref{fig2}.

\section{Nd scattering}
\label{ndscat}

We are now in the position to discuss our predictions for
nucleon-deuteron (Nd) scattering observables. To this aim, we
calculate a 3N scattering operator $T$ by solving the
Faddeev-type integral equation~\cite{wit88,Gloeckle:1995jg,hub97,book} in a
partial wave momentum-space basis. Throughout this section, we
restrict ourselves to the harder regulator value of $R = 0.9$~fm in
order to cover a broader kinematical range up to $E_{\rm lab} =
250$~MeV \footnote{The results
  for low-energy scattering observables using $R=1.0$~fm are comparable
  to the ones using $R=0.9$~fm, see also Ref.~\cite{Binder:2018pgl}
  for a similar conclusion for calculations based on NN forces only. More details will be given in a
  separate publication \cite{fut3ncont}.}  
and focus on
a very restricted set of observables. A more detailed discussion of
Nd elastic and breakup scattering at N$^2$LO will be published elsewhere.  
Since we are going to compare our 3N scattering predictions with
pd data, we have replaced the neutron-neutron (nn) components of the
NN potential with the corresponding proton-proton (pp) ones (with 
the Coulomb force being subtracted).  Further, in order to provide 
converged results, we have solved the 3N Faddeev equations  
by taking into account all partial wave states with the 2N total angular 
momenta up to $j_{max}=5$ and 3N total angular momenta up to $J_{max}=25/2$. 
The 3NF was included up to $J_{max}=7/2$. 

\begin{figure}[tb]   
  \includegraphics[width=0.675\columnwidth]{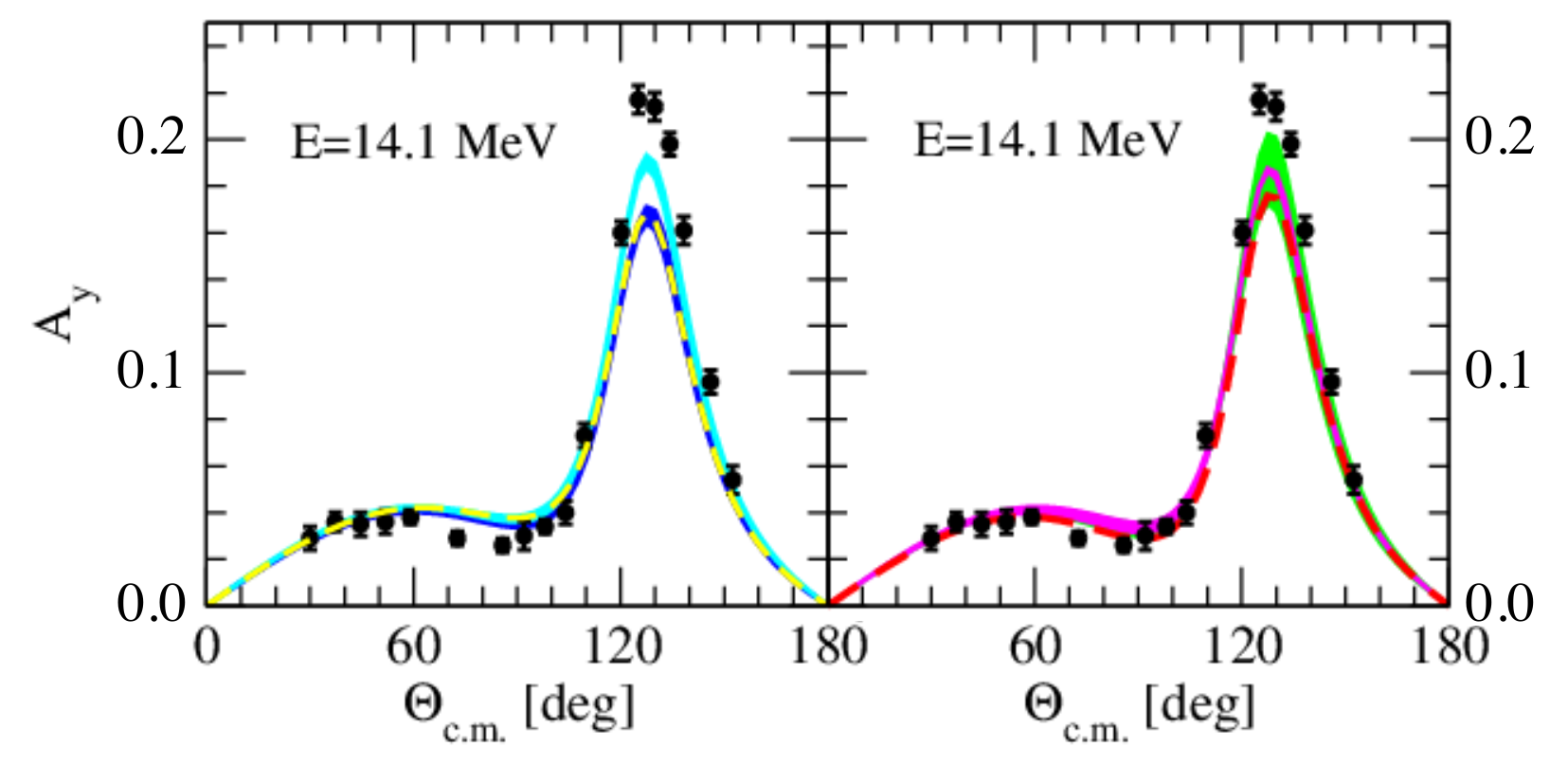}
\caption{(Color online) 
 The neutron analyzing power $A_y$ in nd elastic scattering
 at $E_n=14.1$~MeV. 
The left panel shows the predictions based on the phenomenological NN potentials
AV18, CD~Bonn, Nijm1 and Nijm2
alone (blue band) or in combination with the TM99 3NFs (cyan band). The  dashed (yellow) line is the result 
based on the AV18 NN 
potential in combination with the Urbana IX 3NF. In the right panel,
the dashed (red) line is the prediction of 
the N$^2$LO SCS NN potential with the regulator $R=0.9$~fm. 
 The (magenta) band covers the predictions obtained with 
 this N$^2$LO NN potential combined 
with the N$^2$LO 3NF using $c_D = -2.0 \ldots 6.0$ (and the corresponding 
 $c_E$ values fixed from the correlation line). The (green) band
 gives the estimated theoretical uncertainty  
 at N$^2$LO for the  value of $c_D=2.0$.
 The (black) dots  depict pd data from Ref.~\cite{ay_14.1}.
\label{fig4}}
\end{figure}

At low energies,
 the most interesting observable is the analyzing
  power $A_y$ for nd elastic scattering with polarized neutrons. 
 Theoretical predictions of the phenomenological high-precision NN  
potentials such as the AV18 \cite{AV18}, CDBonn \cite{cdb}, Nijm1 and
Nijm2 \cite{nijm} fail to explain the experimental data for
$A_y$ as visualized in Fig. ~\ref{fig4}.
The data are   underestimated 
by $\approx 30 \%$ in the region of the $A_y$ maximum which
corresponds to  the c.m.~angles of $\Theta_{c.m.} \approx 125 ^\circ$. Combining these NN 
potentials  with the $2\pi$-exchange TM99 3NF model \cite{TM99} removes 
approximately only half of the discrepancy  to the 
data (see Fig.~\ref{fig4}). That effect is, however, model dependent: 
if the Urbana IX 3NF model \cite{uIX} 
is used instead of the TM99 3NF, one observes practically no effects
on $A_y$,  see the left panel of Fig.~\ref{fig4}. The predictions for
the $A_y$ based on the chiral NN potentials appear to be similar to those of
phenomenological models, see \cite{Binder:2018pgl} and references
therein. Combining the N$^2$LO SCS chiral potential  with  
 the N$^2$LO 3NF only slightly  improves
the description of $A_y$. The behavior is qualitatively similar to the
one observed for the TM99 3NF, but the effect is $\sim 2$ times
smaller in magnitude. Interestingly, the theoretical predictions
appear to be quite insensitive to the actual value of $c_D$ as
visualized by a rather narrow magenta band in the right panel of
Fig.~\ref{fig4}, which corresponds to the variation of $c_D = -2.0
\ldots 6.0$. In fact, this observable is well known to be 
very sensitive only to $^3P_j$ NN force components \cite{aypuzzle}, 
while both 3NF contact interactions act predominantly in the S-waves. 
On the other hand, the theoretical uncertainty at N$^2$LO is rather
large and, in fact, comparable in magnitude with the observed deviation between
the predictions and experimental data. It would be interesting to see
whether the $A_y$-puzzle would persist upon inclusion of higher-order
corrections to the 3NF. As for other Nd elastic
scattering observables at low energy, we found the effects of the chiral 3NF at
N$^2$LO to be rather small, and the good description of the data, already
reported in Ref.~\cite{Binder:2015mbz} for the calculations based on the
NN forces, remains intact after inclusion of the 3NF. 

\begin{figure}   
  \includegraphics[width=0.64\columnwidth]{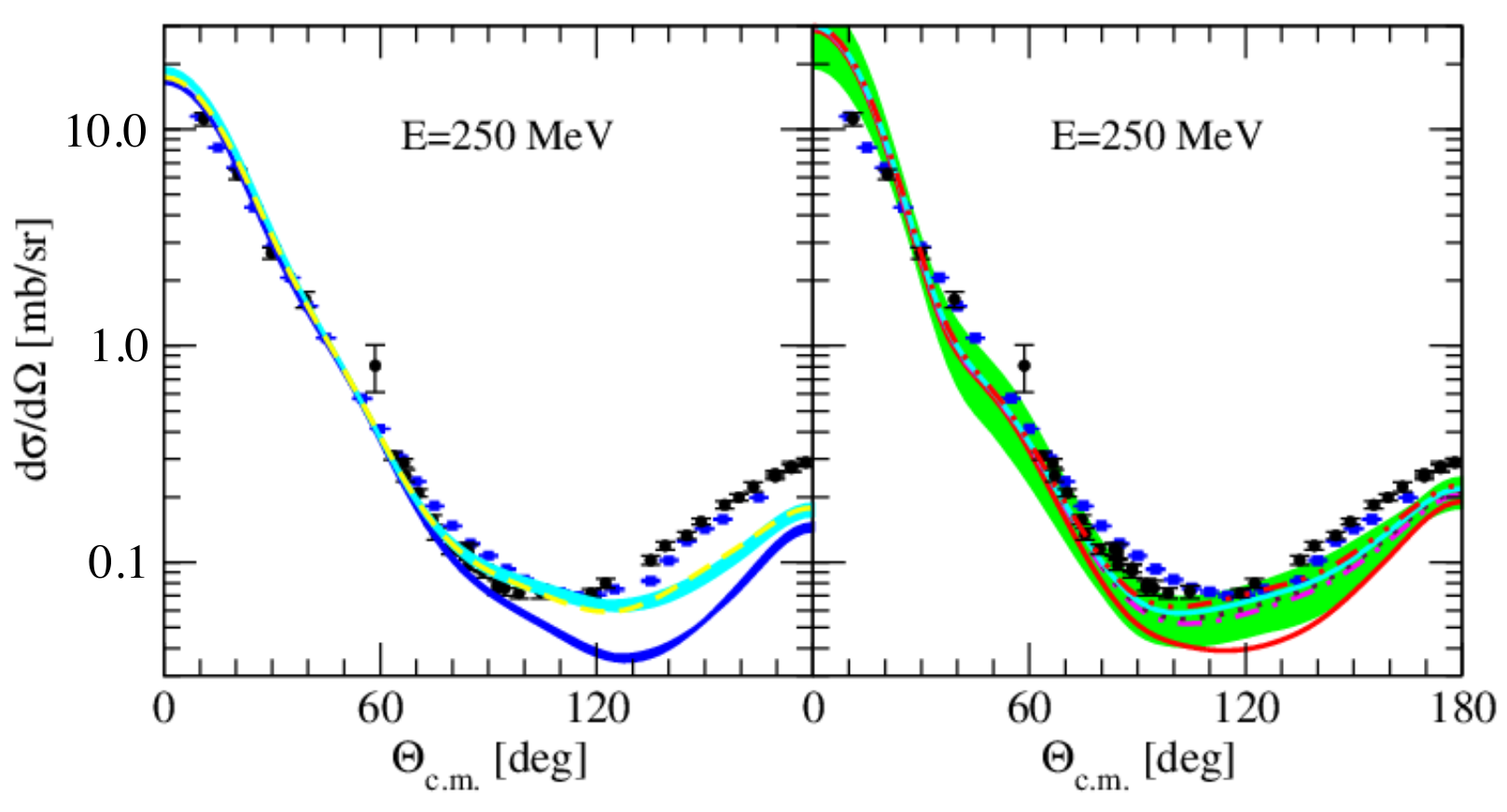}
\caption{(Color online) 
The nd elastic scattering cross section at $E_n=250$~MeV. The lines
and bands in the left (right) panel are the
 same as in the left (left) panel of Fig.~\ref{fig4} (Fig.~\ref{fig2}). 
 (Black) dots depict the pd data from Ref.~\cite{hatanaka02} while 
 (blue) squares are nd data from Ref.~\cite{maeda06}. 
\label{fig3}}
\end{figure}

At intermediate energies, the effects of the 3NF start to become more
pronounced. In particular, as already discussed in section
\ref{cdandce}, the differential cross section is significantly
underestimated in the minimum region when calculated based on NN
forces only. The same pattern is observed in calculations based on the
high-precision phenomenological potentials as well. The improved 
description of Nd elastic scattering cross section data up to 
about $130$~MeV upon inclusion of the N$^2$LO 3NF resembles the
situation found in calculations based on phenomenological 3NFs \cite{wit98,wit2001}
such as  the TM99 \cite{TM99}  and Urbana IX \cite{uIX}
models. On the other hand, the inclusion of the available 3NFs 
has so far not provided an explanation of the  growing discrepancies between
the cross section data and theoretical predictions at larger energies
and backward angles as exemplified in Fig.~\ref{fig3} for $E_N
=250$~MeV. 
The astonishing similarity of the predictions based on
phenomenological models and chiral interactions can presumably be
traced back to the fact that the basic mechanism  
underlying these 3NF's is the $2\pi$-exchange. It is also interesting
to observe that the N$^2$LO  theoretical predictions are rather
insensitive to the variation of $c_D$, $c_E$. Clearly, the convergence
of the chiral expansion at such high energies is expected to be rather
slow as reflected by
the broad error band in the right panel of this figure. In fact, given
the theoretical uncertainty of our N$^2$LO results, the description of the
experimental data appears to be adequate at this chiral order. 

\begin{figure}   
  \includegraphics[width=0.71\columnwidth]{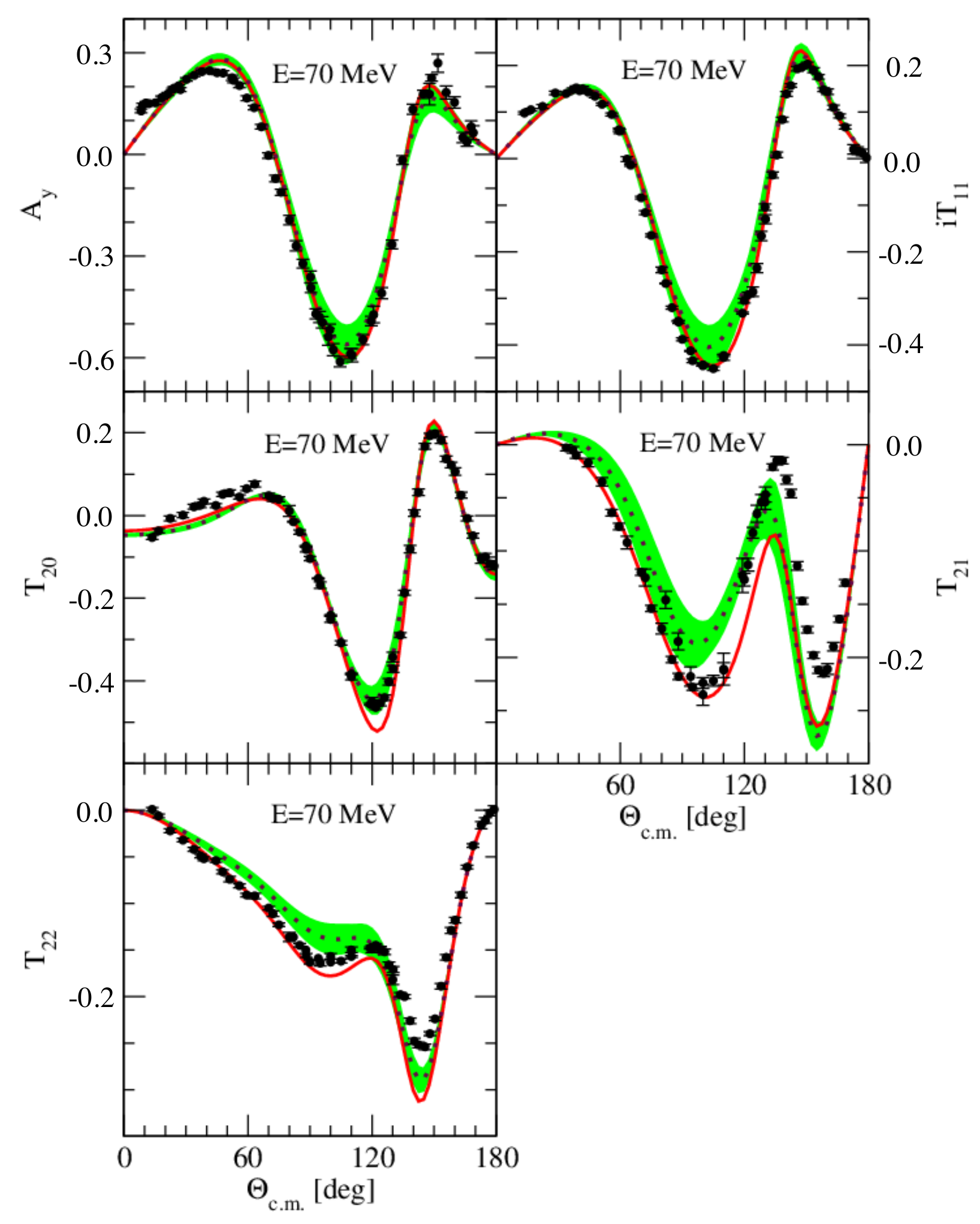}
  \caption{(Color online)  
 The nd elastic scattering neutron ($A_y$) and deuteron ($iT_{11}$) 
 vector analyzing powers 
as well as deuteron tensor analyzing powers $T_{20}$, $T_{21}$ and $T_{22}$ 
at the  incoming neutron laboratory
 energy $E=70$~MeV. The solid red lines 
 are predictions of 
the N$^2$LO SCS NN potential with the regulator $R=0.9$~fm. Combining that 
NN potential with N$^2$LO 3NF with  strengths  of the contact 
terms ($c_D=2.0$,$c_E=-0.3446$) leads to results shown by the 
 dotted maroon  lines with their 
 estimated theoretical uncertainty depicted by 
 the green bands. 
 The black dots depict  pd data for $A_y$  at  $E=65$~MeV 
 from Ref.~\cite{shimizu} and for other analyzing powers
 at $E=70$~MeV from Ref.~\cite{Sekiguchi:2002sf}. 
\label{fig6}}
\end{figure}

Finally, as a representative example, we show in Fig.~\ref{fig6} our
predictions for the complete set of analyzing powers 
at $E=70$~MeV together with the estimated theoretical 
uncertainty. 
Except for the tensor analyzing power $T_{21}$ at backward angles, we
observe a reasonably good description of the data given the
uncertainty of our results. Clearly, one will have
to go to higher chiral orders in order to improve the accuracy of the
calculations and to perform more quantitative tests of the theory. 
Work along these lines is in progress.

\section{Ground state energies for $p$-shell nuclei}
\label{sec:nuclei}

For $p$-shell nuclei, we use No-Core Configuration Interaction (NCCI) methods to solve the many-body
Schr\"odinger equation.  These methods have advanced rapidly in recent
years and one can now accurately solve fundamental problems in nuclear
structure and reaction physics using realistic interactions, see e.g.,
Ref.~\cite{Barrett:2013nh} and references therein.  Here we follow
Refs.~\cite{Maris:2008ax,Maris:2013poa} where, for a given
interaction, we diagonalize the resulting many-body Hamiltonian in a
sequence of truncated harmonic-oscillator (HO) basis spaces.  The goal
is to achieve convergence as indicated by independence of the basis
parameters, but in practice we use extrapolations to estimate the
binding energy in the complete (but infinite-dimensional)
space~\cite{Maris:2008ax,Coon:2012ab,Furnstahl:2012qg,More:2013rma,Wendt:2015nba}.
These NCCI calculations were performed on the Cray XC30 Edison and
Cray XC40 Cori at NERSC and the IBM BG/Q Mira at Argonne National
Laboratory, using the code MFDn~\cite{Maris:2010,Aktulga:2014,Shao:2016}.

In order to improve the convergence behavior of the bound state
calculations we employ the Similarity Renormalization Group
(SRG)~\cite{Glazek:1993rc,Wegner:1994,Bogner:2007rx,Bogner:2009bt}
approach that provides a straightforward and flexible framework for
consistently evolving (softening) the Hamiltonian and other operators,
including three-nucleon
interactions~\cite{Jurgenson:2009qs,Roth:2011ar,Jurgenson:2013yya,Roth:2013fqa}.
In the presence of explicit 3NFs, this additional softening of the
chiral interaction is necessary in order to obtain sufficiently
converged results on current supercomputers for $p$-shell nuclei.  The
flow equation for the three-body system is solved using a HO
Jacobi-coordinate basis~\cite{Roth:2013fqa}.  The SRG evolution and
subsequent transformation to single-particle coordinates were
performed on a single node using an efficient OpenMP parallelized
code.

\begin{figure}
  \includegraphics[width=0.45\columnwidth]{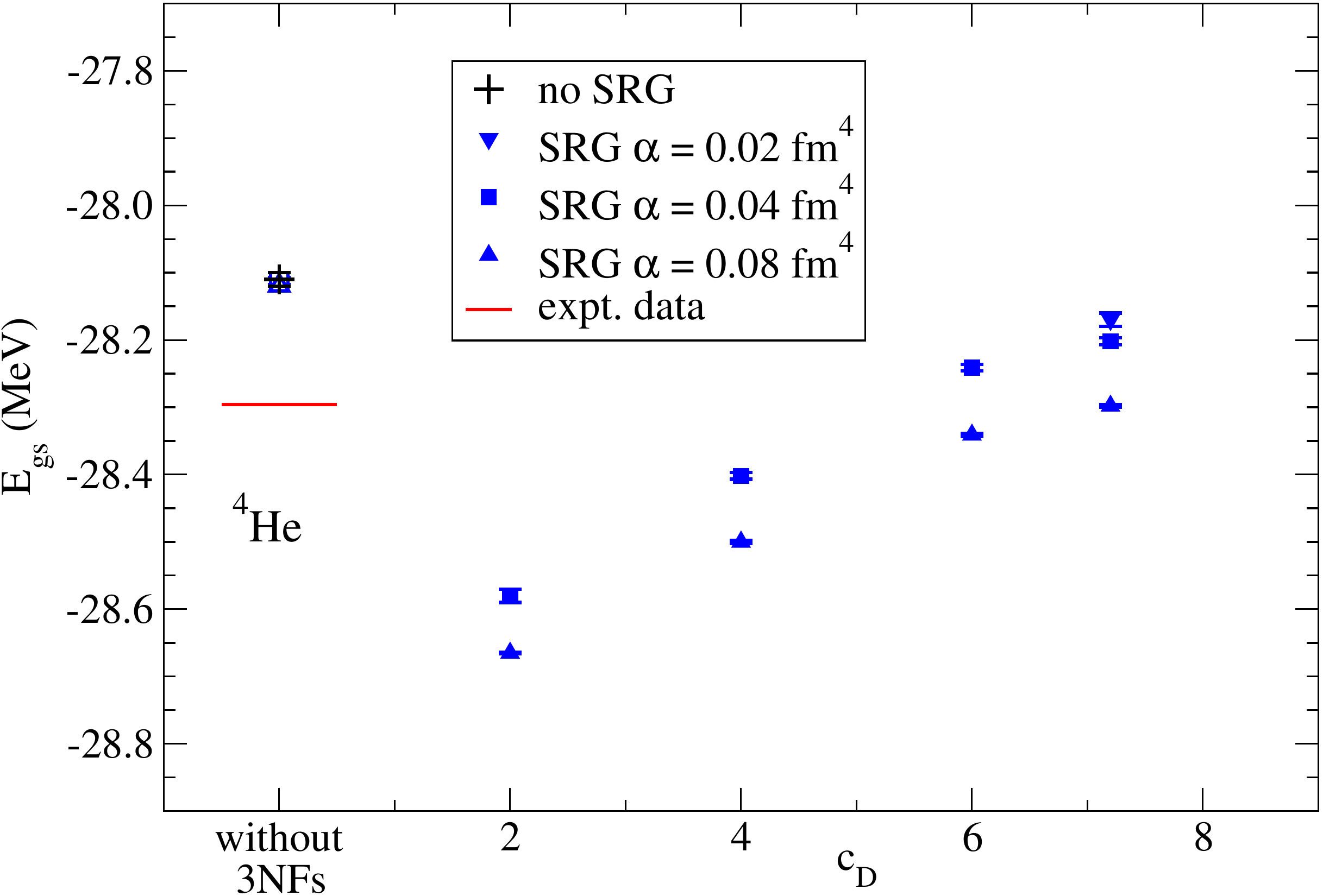}\hfill
  \includegraphics[width=0.44\columnwidth]{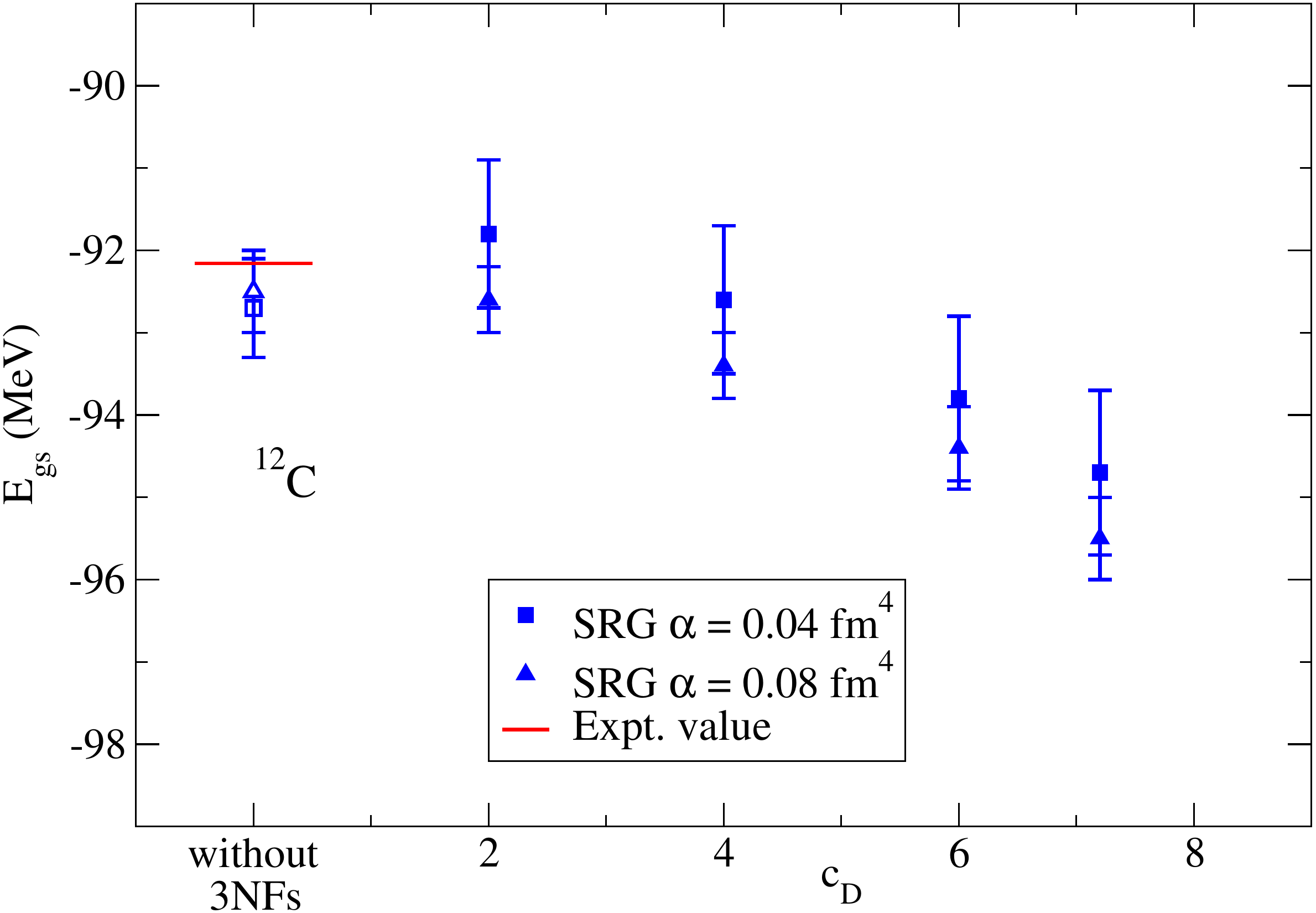}
  \caption{\label{Fig:4He12C}
    Extrapolated ground state energy for $^4$He (left) and $^{12}$C (right)
    using chiral N$^2$LO interactions with regulator $R=1.0$~fm, and
    SRG evolution parameters $\alpha=0.02$, $0.04$, and $0.08$~fm$^4$,
    with and without explicit 3NFs.  The error bars correspond to the
    extrapolation uncertainty estimates only. }
\end{figure} 
As a consequence of the softening of the interaction, our results may
depend on the SRG parameter $\alpha$, because we do not incorporate
any induced interactions beyond 3NFs.  Without explicit 3NFs, this
dependence appears to be negligible, see Fig.~\ref{Fig:4He12C}: for
$^4$He the results with and without SRG evolution are within about
$10$~keV of each other, and for $^{12}$C the difference between the
ground state energies at $\alpha=0.04$ and $\alpha=0.08$~fm$^4$ is
significantly less than the estimated extrapolation uncertainty.
Once we add explicit 3NFs to the NN potential we find that the results
for $^4$He do depend on the SRG parameter, and that this dependence
increases as we evolve the interaction further ($\alpha=0$ corresponds
to the interaction without SRG).  However, for $A \ge 6$ this
dependence becomes of the same order as (or smaller than) our
extrapolation uncertainty estimate.  We can combine the extrapolation
uncertainty and the SRG dependence (estimated by taking the difference
between the binding energies at $\alpha=0.04$ and
$\alpha=0.08$~fm$^4$) into a single numerical uncertainty estimate,
treating them as independent.

In Fig.~\ref{Fig:4He12C} we also see that the binding energies depend
in a nontrivial way on the values of $c_D$ and $c_E$.  In particular,
as we increase $c_D$ (and change the corresponding $c_E$ accordingly)
the ground state energy of $^4$He increases, whereas that of $^{12}$C
decreases with increasing $c_D$.  It turns out that for $A=6$ and $7$
the binding energy is nearly independent (within our numerical
uncertainty estimates) of the actual value of $c_D$, whereas starting
from $A=8$ we do see a systematic decrease of the ground state energy
with increasing $c_D$, at least for $R=1.0$~fm and values of $c_D$
between $2$ and $8$~\cite{Maris:inprep2018}.  Furthermore, this
dependence on $c_D$ seems to be stronger as one moves away from $N=Z$.

\begin{figure}
\includegraphics[width=0.99\columnwidth]{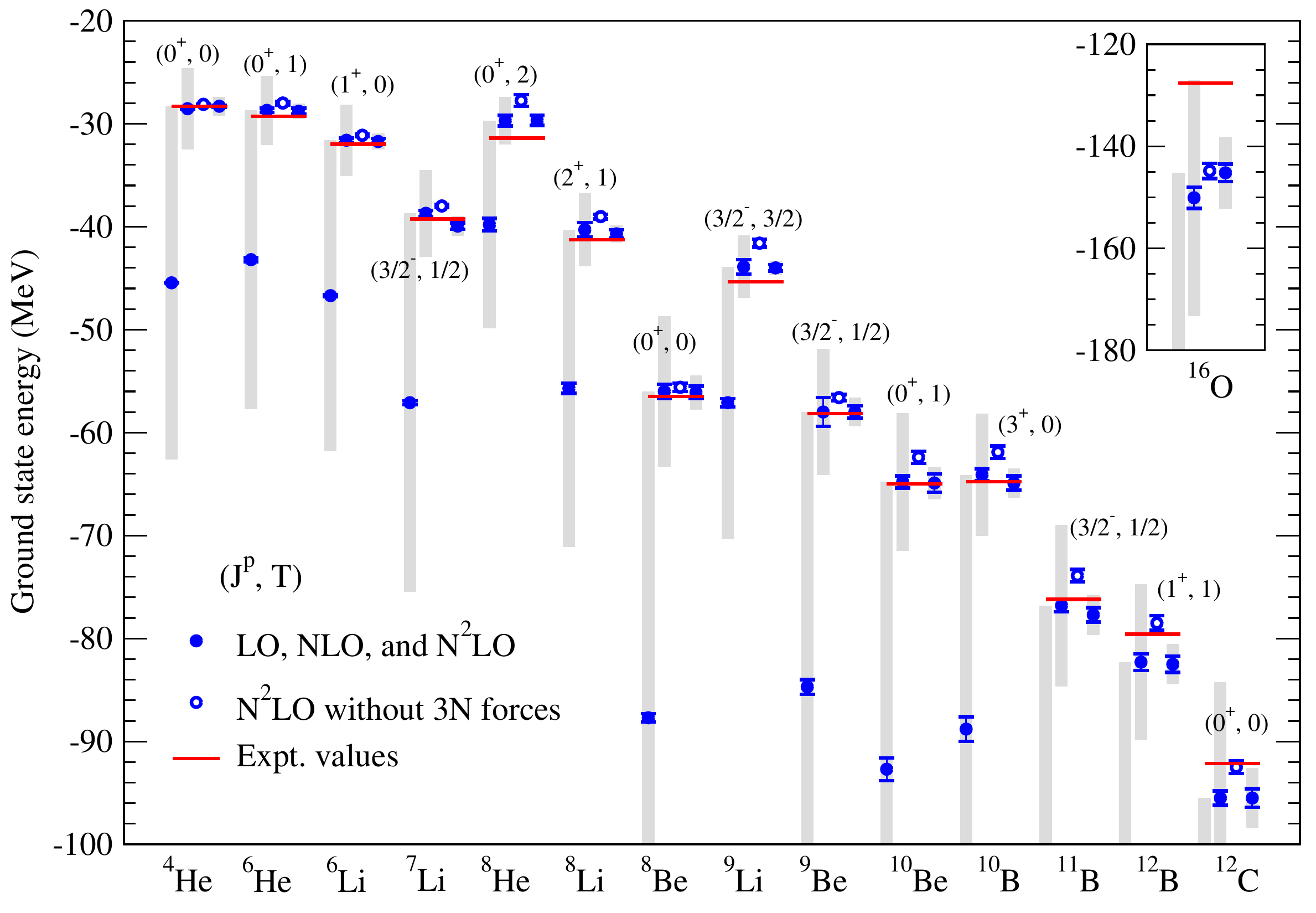}
\caption{\label{Fig:res_gs_Eb}
  (Color online) Calculated ground state energies in MeV using chiral
  LO, NLO, and N$^2$LO interactions at $R=1.0$~fm (blue symbols) in
  comparison with experimental values (red levels).  For each nucleus
  the LO, NLO and N$^2$LO results  are the left, middle and right symbols
  and bars, respectively. The open blue
  symbols correspond to incomplete calculations at N$^2$LO using
  NN-only interactions.  Blue error bars indicate the NCCI
  extrapolation uncertainty and, where applicable, an estimate of the
  SRG dependence.  The shaded bars indicate the estimated truncation
  error at each chiral order following \cite{Binder:2015mbz}.  Note
  that the LO results for $A=11$, $12$, and for $^{16}$O are off the scale,
  but (part of) the corresponding shaded uncertainty bar is included.}
\end{figure} 
We have visualized our results for the ground state energies of $A=4$
to $12$ nuclei in Fig.~\ref{Fig:res_gs_Eb}, for the regulator
of $R=1.0$~fm.  The results at N$^2$LO are all obtained with the
preferred values of $c_D=7.2$ and $c_E=-0.671$ for the LECs,
and an SRG parameter of $\alpha=0.08$~fm$^4$.  The open blue symbols
correspond to {\em incomplete} calculations at N$^2$LO using NN-only
interactions (with induced 3NFs), whereas the complete N$^2$LO
calculations including 3NFs are shown by solid symbols.
For comparison, we have also included the results at LO and NLO
with $R=1.0$~fm.  For $A=4$ through $9$ these calculations at LO
and NLO were performed without SRG evolution~\cite{Binder:2018pgl};
the results for $A=10$, $11$, $12$, and $^{16}$O in
Fig.~\ref{Fig:res_gs_Eb} are for an SRG parameters of
$\alpha=0.08$~fm$^4$, and include induced 3NFs.
(Note that at LO and NLO there are no 3NFs.)

For all $A=4$ to $12$ nuclei the ground state energies decrease when
we add the 3NFs with the preferred LECs to the NN interaction at
N$^2$LO, see Fig.~\ref{Fig:res_gs_Eb}.  For $^4$He this decrease is
very small, but for $A=6$ and larger this decrease is at least half an
MeV, growing to a decrease of about $3$~MeV in the ground state energy
of $^{12}$C.  Up to $A=10$ the ground state energies with the 3NFs are
significantly closer to their experimental values than without;
however, for $A=12$ the decrease of the ground state energies moves
them further away from the experimental value.  In contrast, for
$^{16}$O (see the inset in Fig.~\ref{Fig:res_gs_Eb}) the binding
energy at N$^2$LO is, within the numerical uncertainties, the same
with or without 3NFs, and significantly below the experimental value.

We also show the chiral truncation error estimate for these ground
state energies following
Refs.~\cite{Epelbaum:2014efa,Epelbaum:2014sza,Binder:2015mbz,Binder:2018pgl}.
To be specific, following this method implies that the chiral error
estimate at LO is, in practice, determined by
$\delta E^{(0)} = \max(|E^{(2)}-E^{(0)}|, |E^{(3)}-E^{(0)}|)$,
and at NLO and N$^2$LO by $Q \delta E^{(0)}$ and $Q^2 \delta E^{(0)}$
respectively, where $Q$ is the chiral expansion parameter.  Up to
$A=9$ we use $Q = M_\pi/\Lambda_b\approx 0.23$, but for $A=10$ and
above the average relative momentum scale of the nucleons inside the
nucleus increases, to about $185$~MeV for $^{16}$O, corresponding to
$Q \approx 0.3$~\cite{Binder:2018pgl}.  (It turns out that the chiral
error estimate with 3NFs included at N$^2$LO is up to about 10\%
smaller than those without 3NFs for $A=6$ to $12$.)

For most of the 15 nuclei in Fig.~\ref{Fig:res_gs_Eb}, our complete
results at N$^2$LO agree, to within the chiral error estimate, with
the experimental values; the exceptions are $^8$He, $^9$Li, $^{12}$B,
$^{12}$C, and $^{16}$O.  Both $^8$He and $^9$Li are slightly
underbound in our calculations; they are also both weakly-bound and
neutron-rich.  Small changes in either the two-neutron force or the
three-neutron force (neither of which are very well constrained
experimentally) could potentially have significant effects on these neutron-rich
nuclei.  In this respect it is also interesting to note that the
effect of the 3NFs is noticeably larger for $^8$He and $^9$Li than for
$^8$Be and $^9$Be.  On the other hand, $^{16}$O is noticeably overbound
at N$^2$LO, with or without 3NFs, see also Ref.~\cite{Lahde:2013uqa} for a related
discussion in the context of nuclear lattice simulations.  This overbinding starts at $A=12$,
where, with 3NFs, both $^{12}$B and $^{12}$C are overbound, with the
experimental value only slightly outside the chiral truncation error
estimate, and seems to be systematic for the heavier nuclei.

\begin{table}
  \begin{tabular*}{\textwidth}{@{\extracolsep{\fill}}|l|c|c|c|c|c|c|c|r|}
\hline
     & & & & \multicolumn{2}{c|}{$R = 0.9$~fm} & \multicolumn{2}{c|}{$R = 1.0$~fm} & Exp.
      \\
      Nucleus & $J^P$ & $N_{\max}$ & $\alpha$ [fm$^4$]&
      NN + 3NF$_{\rm induced}$  & NN + 3NF &
      NN + 3NF$_{\rm induced}$ & NN + 3NF  & 
      \\ \hline
      $^4$He   & $0^+$ & 14  & 0.04 &
      $27.231 \pm 0.006$ & $28.425 \pm 0.004$ &
      $28.113 \pm 0.006$ & $28.202 \pm 0.005$ & $28.296$
      \\
               &       &     & 0.08 &
      $27.233 \pm 0.002$ & $28.502 \pm 0.002$ &
      $27.119 \pm 0.001$ & $28.298 \pm 0.002$ &
      \\ \hline
      $^6$He   & $0^+$ & 12  & 0.04 &
      $27.00 \pm 0.16$ & $28.73 \pm 0.15$ &
      $27.88 \pm 0.15$ & $28.55 \pm 0.15$ & $29.27$
      \\
               &       &     & 0.08 &
      $27.10 \pm 0.10$ & $28.94 \pm 0.08$ &
      $27.99 \pm 0.14$ & $28.79 \pm 0.08$ &
      \\ \hline
      $^6$Li   & $1^+$ & 12  & 0.04 & 
      $30.15 \pm 0.15$ & $31.79 \pm 0.18$ &
      $31.02 \pm 0.13$ & $31.49 \pm 0.16$ & $31.99$
      \\
               &       &     & 0.08 &
      $30.24 \pm 0.07$ & $32.00 \pm 0.07$ &
      $31.12 \pm 0.08$ & $31.72 \pm 0.06$ &
      \\ \hline
      $^7$Li   & $\frac{3}{2}^-$ & 10  & 0.04 & 
      $36.89 \pm 0.25$ & $39.04 \pm 0.30$ &
      $37.91 \pm 0.20$ & $38.66 \pm 0.28$ & $39.24$
      \\
               &       &     & 0.08 &
      $36.92 \pm 0.12$ & $39.19 \pm 0.14$ &
      $37.99 \pm 0.11$ & $38.94 \pm 0.14$ &
      \\ \hline
      $^8$He   & $0^+$ & 10  & 0.04 &
      $26.9  \pm 1.0$  & $29.6  \pm 0.5$ &
      $27.5  \pm 0.4$  & $29.3  \pm 0.4$ & $31.41$
      \\
               &       &     & 0.08 &
      $26.87 \pm 0.4$  & $29.88 \pm 0.4$ &
      $27.75 \pm 0.5$  & $29.66 \pm 0.4$ &
      \\ \hline
      $^8$Li   & $2^+$ & 10  & 0.04 &
      $37.87 \pm 0.3$  & $40.85 \pm 0.4$ &
      $38.92 \pm 0.3$  & $40.38 \pm 0.4$ & $41.28$
      \\
               &       &     & 0.08 &
      $37.90 \pm 0.15$ & $41.07 \pm 0.25$ &
      $39.02 \pm 0.2$  & $40.70 \pm 0.2$  &
      \\ \hline
      $^8$Be   & $0^+$ & 10  & 0.04 &
      $53.7  \pm 0.3$  & $56.2  \pm 0.5$  &
      $55.4  \pm 0.4$  & $55.6  \pm 0.5$  & $56.50$
      \\
               &       &     & 0.08 &
      $53.8  \pm 0.2$  & $56.6  \pm 0.3$  &
      $55.6  \pm 0.3$  & $56.1  \pm 0.3$  &
      \\ \hline
      $^9$Li   & $\frac{3}{2}^-$ & 10  & 0.04 &
      $40.5  \pm 0.4$   & $44.1  \pm 0.4$ &
      $41.6  \pm 0.4$  & $43.9  \pm 0.4$ & $45.34$
      \\
               &       &     & 0.08 &
      $40.44 \pm 0.2$   & $44.50 \pm 0.2$ &
      $41.63 \pm 0.3$  & $44.04 \pm 0.2$ &
      \\ \hline
      $^9$Be   & $\frac{3}{2}^-$ & 10  & 0.04 &
      $54.8  \pm 0.4$  & $57.8  \pm 0.5$  &
      $56.5  \pm 0.4$  & $57.5  \pm 0.5$  & $58.16$
      \\
               &       &     & 0.08 &
      $54.81 \pm 0.2$  & $58.42 \pm 0.25$  &
      $56.57 \pm 0.2$  & $58.04 \pm 0.25$  & $58.16$
      \\ \hline
      $^{10}$Be & $0^+$ &  8  & 0.08 &
      $60.4 \pm 0.5$ & $65.6 \pm 0.5$ &
      $62.4 \pm 0.5$ & $64.9 \pm 0.5$ &  $64.98$
      \\
      $^{10}$B  & $3^+$ &  8  & 0.08 &
      $60.0 \pm 0.5$ & $66.0 \pm 0.5$ &
      $61.9 \pm 0.5$ & $64.9 \pm 0.5$ &  $64.75$
      \\
      $^{11}$B  & $\frac{3}{2}^-$ & 8 & 0.08 &
      $71.7 \pm 0.5$ & $78.8 \pm 0.5$ &
      $73.9 \pm 0.5$ & $77.7 \pm 0.5$ & $76.21$
      \\
      $^{12}$B  & $1^+$ &  8  & 0.08 &
      $76.2 \pm 0.5$ & $83.7 \pm 0.6$ &
      $78.5 \pm 0.6$ & $82.5 \pm 0.5$ & $79.58$
      \\
      $^{12}$C  & $0^+$ &  8  & 0.08 &
      $89.7 \pm 0.4$ & $96.9 \pm 0.5 $ &
      $92.5 \pm 0.5$ & $95.5 \pm 0.5 $ & $92.16$ 
      \\ \hline
      $^{16}$O  & $0^+$ &  8  & 0.08 &
      $140.6$ (CR-CC) & $146.9 \pm 0.8$ &
      $144.8 \pm 0.6$ & $145.2 \pm 0.8$ & $127.62$ 
      \\ \hline
  \end{tabular*}
  \caption{\label{Tab:res_gs_Eb}
    Extrapolated binding energies of $A = 6$ to $12$ nuclei in MeV,
    as well as $^{4}$He and $^{16}$O, with the chiral interactions at
    N$^2$LO using semilocal coordinate space regulators, as well as
    SRG evolution to improve numerical convergence of the many-body
    calculations.  For the LEC $c_D$,
    we use the value of $c_D = 2.1$ for $R=0.9$~fm and $c_D = 7.2$ for
    $R=1.0$~fm.  The uncertainty estimate is only the extrapolation
    uncertainty in the many-body calculation, and does not include
    any SRG uncertainty, the chiral truncation error, nor any
    uncertainty due to uncertainties in the LECs. }
\end{table}
Table~\ref{Tab:res_gs_Eb} gives our calculated results at N$^2$LO for
both $R=0.9$~fm and $R=1.0$~fm.  Although the qualitative behavior is
similar for the two regulator values, that is, the explicit 3NFs
at N$^2$LO decrease the ground state energy for all $A=6$ to $12$
nuclei, the additional binding from these 3NFs is significantly larger
at $R=0.9$~fm than at $R=1.0$~fm.  For both regulator values the
additional binding from the 3NFs leads to a better agreement with the
data up to about $A=10$ or $11$.  Furthermore, the regulator dependence is noticeably smaller
with the 3NFs included than without these contributions.

However, inclusion of the 3NFs leads to a noticeably overbinding for
both $^{12}$B and $^{12}$C, whereas the effect of the 3NFs is
surprisingly small for $^{16}$O, and does not move the $^{16}$O
binding energy any closer to experiment.  Note that a smaller value of
$c_D$ would give better agreement with the experimental binding energy
for $^{12}$C: with a values of $c_D=2.0$ and $c_E=-0.193$ for the LECs
using $R=1.0$~fm,
the ground state energy of $^{12}$C is in perfect agreement with its
experimental value, see Fig~\ref{Fig:4He12C}.

It is interesting to compare our results to similar calculations using
different versions of the chiral interactions. Our N$^2$LO results for the
ground state energies of $p$-shell nuclei are in a qualitative
agreement with the Green's function Monte Carlo calculations reported in
\cite{Piarulli:2017dwd} and based on the local NN
potentials with explicit $\Delta$ contributions to the two-pion
exchange, accompanied with the locally regularized 3NF at N$^2$LO. 
In particular,  the ground state energy of $^{12}$C, the heaviest
nucleus considered in that work, appears to be slightly overbound at
N$^2$LO. It is, however, difficult to make a more quantitative
comparison since the authors of that paper do not show results  at
lower chiral orders and at N$^2$LO using NN interactions only. Also
no estimation of the theoretical uncertainty is provided. Another
local version of the chiral NN interaction, constructed in
Refs.~\cite{Gezerlis:2013ipa,Gezerlis:2014zia} and 
accompanied with the locally regularized 3NF at N$^2$LO,
was employed in Refs.~\cite{Lonardoni:2017hgs,Lonardoni:2018nob} to
calculate properties of nuclei up to 
$A=16$ using the auxiliary field diffusion Monte Carlo methods. 
This interaction leads to similar results at LO, showing typically a
strong overbinding for all nuclei. The NLO local forces used in
Refs.~\cite{Lonardoni:2017hgs,Lonardoni:2018nob}, however, turn out to be considerably more repulsive than the 
semilocal interactions employed in our analysis, which results in
underbinding for most of the considered nuclei. Still, their NLO
results are consistent with ours and with experimental data within
errors. At N$^2$LO, the authors of Refs.~\cite{Lonardoni:2017hgs,Lonardoni:2018nob} do
not provide results based on the NN interactions only, leaving no
possibility to quantify 3NF effects in their scheme. It is furthermore found in
Ref.~\cite{Lonardoni:2017hgs} that, while being equivalent modulo higher-order terms,
different operator choices of the contact 3NF 
at N$^2$LO may induce large differences for the $^{16}$O binding energy
for (very) soft cutoff values. This indicates that subleading
short-range 3NF contributions may play an important role, especially for
soft choices of the regulator. The description of the ground state
energies of nuclei up to $A=16$ reported in Ref.~\cite{Lonardoni:2018nob} at N$^2$LO is
comparable to ours, but the results for $^{12}$C and $^{16}$O show the
opposite trend of being underbound. We further emphasize that the
short-range part of the 3NF was constrained in that paper in the $A=5$
system using experimental data on $n$-$\alpha$ scattering, while all
results of our calculations for $A \ge 4$ are  parameter-free predictions.

\section{Excitation energies for $p$-shell nuclei}
\label{sec:nuclei2}

\begin{figure}
\includegraphics[width=0.99\columnwidth]{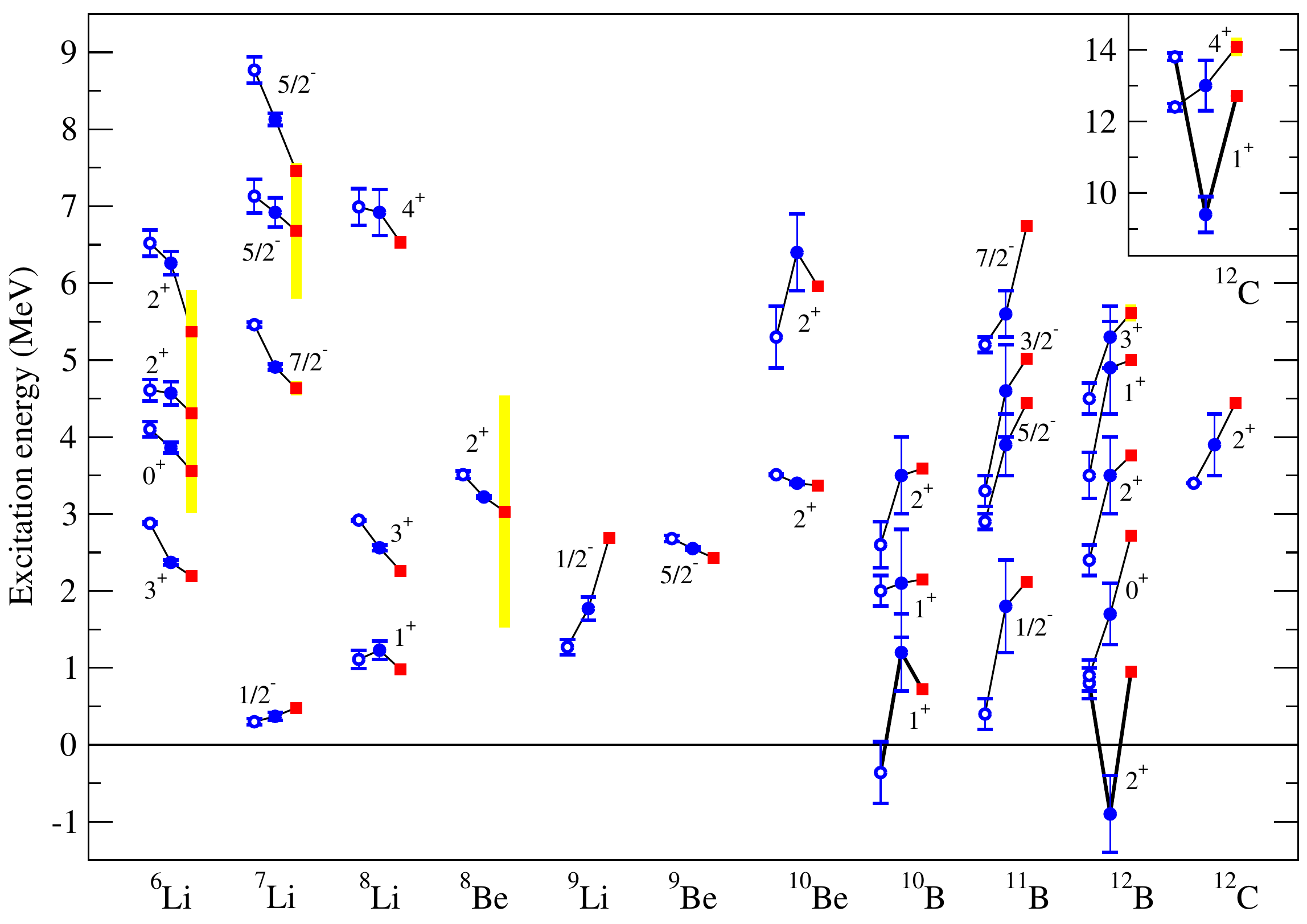}
\caption{\label{Fig:res_Ex_A06A12}
  (Color online) Calculated excitation energies in MeV using chiral
  N$^2$LO at $R=1.0$~fm with and without explicit 3NFs for a fixed
  SRG parameter of $\alpha=0.08$~fm$^4$ and fixed HO basis parameter
  of $\hbar\omega=20$~MeV.
  Results are presented as open blue circles for calculations without explicit 3NFs,
  solid blue dots for calculations including 3NFs using $c_D=7.2$ and 
  red squares for experimental values.  The yellow bars are the
  experimental width of broad resonances.  We define an 'uncertainy
  range' for our calculations by the maximum of the difference between
  calculations at $\hbar\omega=20$~MeV and those at
  $\hbar\omega=16$~MeV or $\hbar\omega=24$~MeV.}
\end{figure} 
In Fig.~\ref{Fig:res_Ex_A06A12} we present our results for the
excitation energies for selected states of $A = 6$ to $12$ nuclei, at
N$^2$LO with $R = 1.0$~fm, first without explicit 3NFs (open blue
symbols), then with explicit 3NFs using the preferred values of
$c_D=7.2$, $c_E=-0.671$ for the LECs (closed blue symbols), and
followed by the experimental values.  All of the shown results were
obtained in the largest achievable basis space in the $N_{\max}$
truncation, and for a fixed SRG parameter of $\alpha=0.08$~fm$^4$ and
fixed HO basis parameter of $\hbar\omega=20$~MeV.  We include the
maximum of the difference between our results at $\hbar\omega=20$~MeV
and those at $\hbar\omega=16$~MeV or $\hbar\omega=24$~MeV as a rough
estimate of the numerical uncertainty of our calculations. 

The results clearly show that including the 3NFs move the excitation
energies for most of these states closer to the experimental values.
There are only two significant exceptions, both for $A=12$: the lowest
$2^+$ state of $^{12}$B and the lowest $1^+$ state of $^{12}$C.  Both
of these two states are in better agreement with experiment without
3NFs than with the 3NFs, and for both, including the 3NFs lowers
the excitation energies significantly.

In $^{12}$B we actually find that the lowest $2^+$ state becomes the
ground state when the 3NFs are included, at almost one MeV below the
actual $1^+$ ground state.  From Fig.~\ref{Fig:res_gs_Eb} we can see
that also the $1^+$ ground state becomes deeper bound when the 3NFs
are included, but the additional binding from the 3NFs is apparently
stronger for the lowest $2^+$ state than for the lowest $1^+$ state.
In contrast, the excitation energies of the other excited states of
$^{12}$B increase when the 3NFs are included, and move considerably
closer to their experimental values.

The excitation energy of the lowest $1^+$ state of $^{12}$C (with $T=0$; the
analog state of the ground state of $^{12}$B is around $15$~MeV, in
agreement with experiment) drops by about $4$~MeV when the 3NFs are
included, from about $14$~MeV to below $10$~MeV, whereas the
experimental value is at $12.7$~MeV.  We find a similar dependence of
this state on the 3NFs using the regulator $R=0.9$~fm, and also with
the Entem--Machleidt chiral N$^3$LO NN potential plus N$^2$LO
3NFs~\cite{Maris:2014hga,Calci:2016tvb}.  Of course, our calculations
are not converged, and in particular for $^{12}$C it is known that the
first excited $0^+$ state (the Hoyle state) cannot be represented in
the finite HO bases that we are employing in our calculations, and is
indeed absent from the low-lying spectrum in our calculations in basis
spaces up to $N_{\max}=10$.  It is possible that this $1^+$ state
is also sensitive to configurations that are beyond $N_{\max}=10$, whereas
the $2^+$ and $4^+$ excited states are rotational excitations of the
ground state and having a similar structure as the ground
state and, therefore, converge at similar rates as the ground state.

In the case of $^{10}B$ we find the now-accepted result of obtaining a
$1^+$ ground state without 3NFs~\cite{Caurier:2002hb} instead of the
observed $3^+$ ground state.  When we include consistent 3NFs, we do
obtain a $3^+$ ground state in concert with experiment, as may be
expected~\cite{Navratil:2007we}.  The excitation energies of the two
additional $^{10}B$ states shown in Fig.~\ref{Fig:res_Ex_A06A12}, the
second $1^+$ state and a $2^+$ state, move closer to experiment with
the addition of the 3NFs as well.  However, the two low-lying $1^+$
states exhibit a strong mixing~\cite{Jurgenson:2013yya}, which results
in a large basis space dependence for these two states, as well as
sensitivity to the SRG parameter, preventing us from reliably
extracting their excitation energies.

Although these three states are sensitive to the LECs $c_D$ and $c_E$,
we do find a qualitatively similar effect if we change $c_D$ over a
range from $2$ to $8$~\cite{Maris:inprep2018}.  
However, a lower value of $c_D$ would improve the agreement with
experiment for the $1^+$ state of $^{12}$C and the $2^+$ state of
$^{12}$B: with $c_D=2.0$ and $c_E=-0.193$ the $2^+$ state of $^{12}$B
becomes essentially degenerate with the $1^+$ ground state, and the
excitation energy of the $1^+$ state of $^{12}$C becomes about
$11.2$~MeV, that is, significantly closer to the experimental value.
For $^{10}$B the situation is much more complicated, due to the strong
mixing between the lowest two $1^+$ states as a function of the basis
truncation parameters $N_{\max}$ and
$\hbar\omega$~\cite{Jurgenson:2013yya}.  Note however that none of
these excitation energies are very well converged.  The excitation
energies of most of the other states shown here are significantly less
sensitive to the LECs.

Another interesting observation is that for $A=6$, $7$, and $8$ the
inclusion of the 3NFs tends to reduce the excitation energies, whereas
for $A=10$, $11$, and $12$ the inclusion of the 3NFs tends to increase
the excitation energies (with the exception of the aforementioned
three states).  Furthermore, both tendencies move the excitation
energies closer to their experimental values.  Nevertheless, even with
the 3NFs included, the calculated excitation energies tend to be too
large for $A=6$, $7$, and $8$ (i.e. the spectrum is too spread out),
whereas for $A=11$ and $12$ they tend to be too small (i.e. the
spectrum is too compressed).  

\section{Summary and outlook}
\label{secsummary}

In this paper we applied the SCS N$^2$LO chiral NN potential 
combined with the N$^2$LO 3NF, regularized in the same way, to
selected properties of few- and many-nucleon systems up to $A=16$. The
main findings of our study can be summarized as follows: 
\begin{itemize}
\item
We have explored the possibility to determine the LECs $c_D$ and $c_E$
from a range of observables in the 3N system. To this aim we first
computed numerically the LECs $c_E$ as a 
function of $c_D$ from the requirement that the $^3$H binding energy
is correctly reproduced.  To fix the value of $c_D$ we have calculated
the Nd doublet scattering length as well as the differential and total
cross sections in Nd scattering at various energies.
By taking into account the estimated truncation
error at N$^2$LO, we found the Nd doublet scattering length to yield
only very weak constraints on the allowed $c_D$ values. These findings
support the conclusions of Ref.~\cite{Gazit:2008ma} and can be traced back
to the strong correlation between this observable and the $^3$H
binding energy known as the Phillips line \cite{Phillips:1968zze}. 
From the considered 3N observables, the strongest constraint
on the $c_D$ values is found to emerge from  the precise experimental data of
Ref.~\cite{Sekiguchi:2002sf} for the differential 
cross section at $E_N = 70$~MeV in its minimum region. The
constraints on the LEC $c_D$ placed by all considered observables appear
to be mutually consistent within errors with the only exception 
of the total cross section at $E_{N} = 135$~MeV for the softer cutoff
of $R=1.0$~fm.  A global analysis of all considered scattering
observables is shown to allow for a precise determination of the LEC
$c_D$ for both considered cutoff values. 
\item
The resulting nuclear Hamiltonian at N$^2$LO has been applied to a
selected range of other observables in elastic Nd scattering. For the low-energy 
nucleon analyzing power $A_y$, the application of consistent 
chiral interactions supports earlier findings based 
on the phenomenological NN potentials accompanied by the TM99 3NF, 
however, the resulting effects are smaller in magnitude by a factor of $\sim 2$.  
We have also looked at various spin observables at $E_N = 70$~MeV,
which turn out to be reasonably well described given the estimated
theoretical uncertainty at this order. At higher energies the
discrepancies between the calculated observables and experimental data
increase, but it is difficult to draw definite conclusions due to rather
large truncation errors at this chiral order. 
\item
Using NCCI methods, we have studied the ground state and low-lying
excitation energies of $p$-shell nuclei. For almost all considered
cases with very few exceptions such as e.g.~the $A=12$ nuclei,  adding the
consistent 3NF to the NN interaction  is found to significantly
improve the description of experimental data. The predicted ground
state energies of $p$-shell nuclei show a good agreement with the data
except for $^{16}$O, which appears to be overbound. 
\end{itemize}
To summarize, we obtain very promising results for a broad range of
few- and many-nucleon observables at N$^2$LO of the chiral expansion. 
In the future, we plan to extend these studies beyond this chiral
order
\cite{Epelbaum:2006eu,Epelbaum:2007us,Ishikawa:2007zz,Bernard:2007sp,Bernard:2011zr,Girlanda:2011fh,Krebs:2012yv,Krebs:2013kha,Epelbaum:2014sea},
see Refs.~\cite{Golak:2014ksa,Hebeler:2015wxa} for first steps along these lines, 
which will allow us to improve the accuracy of our predictions
and perform more stringent tests of the theoretical framework. 
Notice, however, that the coordinate-space regularization of the 3NF
and its subsequent partial wave decomposition represent highly nontrivial tasks
starting from N$^3$LO. Fortunately, this major obstacle can now be
overcome thanks to the newest momentum-space version of the local
regulator employed in the currently most precise version of the chiral NN potentials of 
Ref.~\cite{Reinert:2017usi}. Work along these lines is in progress.

\begin{acknowledgments}
This study has been performed within Low Energy Nuclear Physics
International Collaboration (LENPIC) project and 
was  supported by BMBF (contracts No.~05P2015 - NUSTAR R\&D and No. 05P15RDFN1 - NUSTAR.DA),
by the European
Community-Research Infrastructure Integrating Activity ``Study of
Strongly Interacting Matter'' (acronym HadronPhysics3,
Grant Agreement n. 283286) under the Seventh Framework Programme of EU,
the ERC project 307986 STRONGINT, by the DFG (SFB 1245), by DFG and
 NSFC (CRC 110), by the Polish National Science Centre 
under Grants No. 2016/22/M/ST2/00173 and 2016/21/D/ST2/01120, 
by the Chinese Academy of Sciences (CAS) Presidents
International Fellowship Initiative (PIFI) (Grant No. 2018DM0034) and
by the US Department of Energy (DOE) under Grant Nos. DE-FG02-87ER40371 and
DE-SC0018223 (SciDAC-4/NUCLEI). 
Numerical
calculations were performed on the supercomputer cluster of the JSC,
J\"ulich, Germany.  Numerical calculations
were also performed at the Argonne Leadership Computing Facility, which is a DOE Office of Science User
Facility supported under Contract DE-AC02-06CH11357 with resources provided by an INCITE award,
Nuclear Structure and Nuclear Reactions, from the US DOE Office of Advanced Scientific Computing.
This research also used computational resources provided by the National Energy Research Scientific
Computing Center (NERSC), which is supported by the US DOE Office of
Science.
\end{acknowledgments}


\end{document}